\documentstyle[12pt]{article}
\begin{document}
\title{CP violation in the heavy neutrinos production process $e^+e^-
\rightarrow N_1N_2$
 \thanks{This work was
supported by Polish Committee for Scientific Researches under Grants
Nos. 2252/2/91 and 2P30225206/93}}
\author{J.Gluza$^{\dag}$ and M.Zra{\l}ek$^{\ddagger}$ \\
Department of Field Theory and Particle Physics \\
Institute of Physics, University of Silesia \\
Uniwersytecka 4,PL-40-007 Katowice, Poland}
\maketitle
PACS number(s): 13.15.-ef,12.15.Cc,14.60.Gh,11.30.Er
\begin{abstract}

The problem of CP conservation and CP violation for two heavy neutrinos
production in $e^+e^-$ interaction is considered. Very convenient
way of parametrization of the neutrino mass matrix, from which necessary and
sufficient condition for CP conservation easily follows, is presented.
Contrary to the Kobayashi-Maskawa mechanism, the effects of CP violation in
the lepton sector with Majorana neutrinos can be very large. Change
of the total cross section caused by CP violation can be
much larger then the cross section itself.
\baselineskip 6mm
\end{abstract}
\section{Introduction}
\ \ \
The origin of CP violation is one of the most important open problems
in particle physics. In the standard model (SM) the CP violation is
explained by the Kobayashi-Maskawa mechanism [1]. In this mechanism
the CP violation depends on mixing between flavour
eigenstates and mass eigenstates.
For the mixing to take place, the fermions with given charges must have
distinguishable
masses. That is why the CP violation is visible in
the quark sector (quark masses are distinguishable) and not visible in the
lepton sector (light neutrinos masses are still consistent with zero).
The CP violation effect has been observed untill now only in $K^0-\bar{K}^0$
sector
[2] and is small. It is because the only quantity which describes the CP
violation
in the KM mechanism is the parameter $\delta_{KM}$ given by
\begin{equation}
\delta_{KM}= Im \left( V_{cd}V_{ub}V_{cb}^{\ast}V_{ud}^{\ast}\right) .
\end{equation}
As the KM mixing matrix parameters $V_{ik}$ are small the $\delta_{KM}$ is also
small
\begin{equation}
\delta_{KM}< {10}^{-4}.
\end{equation}
The CP violation problem is very interesting in the lepton sector if the
neutrinos
are Majorana particles. First of all, contrary to the Dirac particles, the
physical Majorana fields are not rephasing invariant. Then not so much phases
can be eliminated and CP is violated already for two generations of leptons
[3]. The greater number of non-eliminated phase parameters is also the cause
why the
CP violation is not mass suppressed [4] so the effect could be potentially
visible
even for very light neutrinos.

In this paper we consider the problem of CP violation in the case of heavy
Majorana neutrinos.
Such particles with the masses greater then 100 GeV can be produced in the
future $e^+e^-$ colliders.
All our considerations are done in the framework of the Left-Right (L-R)
symmetric model which predicts the existence of the Majorana neutrino in a
natural way.
In the next Chapter we find the most convenient parametrization of the mass
matrix for the study of CP violation. A necessary and sufficient condition
guaranteeing CP invariance on the level of weak lepton states is studied.
The numerical analysis of the CP violation in the $e^+e^- \rightarrow
N_1N_2$ process are done in Chapter 3 and some conclusions are presented at the
end.

\section{Parametrization of the mass and mixing matrices}
\ \ \
We consider the L-R model [5] described in details in Refs[6]. The relevant
parts of the model's lagrangian for studying the CP properties are the
charged-current
interaction and the lepton mass lagrangian.
They are given by
\begin{equation}
L_{CC}=\frac{g}{\sqrt{2}} \left(
\bar{\nu}_L\gamma^{\mu}l_LW_{L\mu}^++\bar{\nu}_R
\gamma^{\mu}l_RW_{R\mu}^+ \right) + h.c.
\end{equation}
and
\begin{equation}
L_{mass}=-\frac{1}{2} \left( \bar{n}_L^cM_{\nu}n_R+\bar{n}_RM_{\nu}^{\ast}
n_L^c \right) - \left( \bar{l}_LM_ll_R+\bar{l}_RM_l^+l_L \right)
\end{equation}
where $n_R$ is six-dimensional vector of the neutrino fields
\begin{eqnarray}
n_R&=&\left( \matrix{ \nu_R^c \cr
                      \nu_R } \right)\;,\; \nu_R^c=i\gamma^2\nu_L^{\ast},
     \nonumber \\
n_L&=&\left( \matrix{ \nu_L \cr
                      \nu_L^c } \right)\;,\; \nu_L^c=i\gamma^2\nu_R^{\ast}.
\end{eqnarray}
$M_{\nu}$ and $M_l$ are $6 \times 6$ and $3 \times 3$ mass matrices for
neutrinos and charged leptons respectively. We consider the model with
the explicite     left-right symmetry where the left-handed neutral Higgs
triplet
does not condensate $(v_L=0)$. Then the mass matrix $M_{\nu}$ is given by
\begin{equation}
M_{\nu}=\left( \matrix{ 0 & M_D \cr
                       M_D^T & M_R } \right)
\end{equation}
where $3 \times 3$ matrices $M_D$ ( and also $M_l$) are hermitian and $M_R$ is
symmetric.
The most general CP transformation which leaves the gauge interactions (3)
invariant is [7]
\begin{eqnarray}
l_L &\rightarrow & V_L Cl_L^{\ast}\;,\;\;\nu_L \rightarrow V_L C\nu_L^{\ast},
\nonumber \\
l_R &\rightarrow & V_R Cl_R^{\ast}\;,\;\;\nu_R \rightarrow V_R C\nu_R^{\ast}.
\end{eqnarray}
where $V_{L,R}$ are $3 \times 3$ unitary matrices acting in
lepton flavour space and C is the Dirac charge conjugation matrix.
For the full lagrangian to be invariant under (7) the lepton mass matrices
$M_D,M_R$ and $M_l$ have to satisfy the conditions
\begin{eqnarray}
V_L^{\dag}M_DV_R&=&M_D^{\ast}, \nonumber \\
V_R^TM_RV_R&=&M_R^{\ast},
\end{eqnarray}
and
\begin{eqnarray}
V_L^{\dag}M_lV_R&=&M_l^{\ast} .
\end{eqnarray}
The relations expressed by Eqs.(8) and (9) are weak-basis independent and
constitute
necessary and sufficient condition for CP invariance. It means that if for
given matrices $M_D,M_R$ and $M_l $, there exist two unitary matrices $V_L$ and
$V_R$ such that relations (8,9) hold then our model is CP invariant
and, on the other hand, if CP is the symmetry of our model then such matrices
$V_L$ and
$V_R$ exist.
The most convenient basis for studying CP symmetry is the weak basis in which
charged lepton mass matrix $M_l$ is real, positive and diagonal
\begin{equation}
M_l=diag[m_e,m_{\mu},m_{\tau}].
\end{equation}
Then for non-degenerate, non-vanishing $m_e \ne m_{\mu} \ne m_{\tau}$ Eq.(8)
and (9)
imply that matrices $V_{L,R}$ are diagonal and equal
\begin{equation}
V_L=V_R=diag[e^{i\delta_1},e^{i\delta_2},e^{i\delta_3}].
\end{equation}
{}From Eqs.(8) and (9) follows that the model has CP symmetry if and only if
the
matrices $M_D$ and $M_R$ have the elements
\begin{eqnarray}
(M_D)_{ij}&=&\mid (M_D)_{ij} \mid e^{+\frac{i}{2}(\delta_i-\delta_j)},
\nonumber \\
(M_R)_{ij}&=&\mid (M_R)_{ij} \mid e^{-\frac{i}{2}(\delta_i+\delta_j)}
\end{eqnarray}
in the basis where $M_l$ is diagonal. The number of reduced phases ($\frac
{n(n+1)}{2}$ for symmetric $M_R$ and $\frac{n(n-1)}{2}$ for hermitian
$M_D$ give totally $n^2$ phases)
\begin{equation}
n^2-n \;\;(=6)
\end{equation}
is the lepton sector number of independent CP violating phases in the
considered model (with explicite L-R symmetry and $v_L=0$).

It is easy to understand why relations (12) are necessary and sufficient
conditions for CP invariance. From Eqs.(12) follows that the neutrino mass
matrix $M_{\nu}$ (Eq.6) is diagonalized by the orthogonal transformation
\begin{equation}
U^TM_{\nu}U=diag[\mid m_1'\mid,...,\mid  m_6'\mid ]
\end{equation}
and the $(2n \times 2n)$ unitary matrix U can be expressed in the form
\begin{equation}
U=\left( \matrix{ V^{\ast} & 0 \cr
                  0 & V } \right) O \eta
\end{equation}
where
\begin{equation}
V=diag[e^{i\delta_1/2},e^{i\delta_2/2},e^{i\delta_3/2}],
\end{equation}
O is a real orthogonal $2n \times 2n$ matrix $(O^T=O^{-1})$ that diagonalizes
the real part of $M_{\nu}$ matrix after removing the phases $e^{i\delta_i/2}$,
and $\eta$ is a diagonal $(2n \times 2n)$ matrix that ensures that the neutrino
masses are positive numbers $(m_i=\mid m_i' \mid \geq 0)$
\begin{equation}
\eta_{ij}=\delta_{ij}e^{i\frac{\pi}{4}(sign[m_i']-1)}.
\end{equation}
The CP symmetry is then satisfied if we define the CP parity of Majorana
neutrinos [8]
\begin{equation}
\eta_{CP}(i)=i sign[m_i'].
\end{equation}
To find the mixing matrices $K_{L,R}$ for the left (right) charged current and
the neutral currents $\Omega_{L,R}$ (see Ref.[6] for precise definition) we
define
\begin{equation}
U\equiv \left( \matrix{ U_L^{\ast} \cr
                 U_R } \right) =
\left( \matrix{ V^{\ast}O_L\eta \cr
                 VO_R\eta } \right).
\end{equation}
Then
\begin{eqnarray}
K_L\equiv U_L^{\dag} &=& \eta O_L^TV^{\dag} \nonumber \\
K_R\equiv U_R^{\dag} &=& \eta^{\ast} O_R^TV^{\dag},
\end{eqnarray}
and
\begin{eqnarray}
\Omega_L \equiv K_LK_L^{\dag} &=&\eta O_L^TO_L\eta^{\ast} ,\nonumber \\
\Omega_R \equiv K_RK_R^{\dag} &=&\eta^{\ast} O_R^TO_R\eta ,\nonumber \\
\Omega_{RL} \equiv K_RK_L^{\dag} &=&\eta^{\ast} O_R^TO_L\eta^{\ast} .
\end{eqnarray}
{}From Eqs.(20) and (21) we see that the phase factors from matrix V
multiply the columns of the matrices $K_{L,R}$ and can be absorbed by
rephasing of the charged-lepton fields in the charged currents
$l_{L,Ri} \rightarrow e^{i\delta_i/2}l_{L,Ri}$.
The phase factors disappear from matrices $\Omega_{L,R}$ and $\Omega_{RL}$
which mix the physical Majorana neutrino fields for which the rephasing
is not possible.\footnote{We adopt the definition of the physical Majorana
fields
N(x) as fields that under charge conjugation stay the same without any
 phase factor
 $$N^C(x)\equiv C \bar{N}^T(x)=N(x).$$

 For definition of Majorana fields where the creation phase factors are
 introduced see Refs.[9]. We do not think that these definitions are useful.}
Then, if the CP is not spontaneously broken, the total lepton lagrangian
(gauge-gauge, gauge-leptons, Higgs-leptons and Higgs interactions) is CP
invariant.
 If the phases of matrices $M_D$ and $M_R$ differ from those that are given
by Eqs.(12) the CP symmetry is broken. In the next Chapter we investigate the
 effect of these CP broken phases in the production process of two heavy
 neutrinos.

 \section{The CP effect in the process $e^+e^- \rightarrow N_1N_2$;
  numerical analysis.}
\ \ \
 The amplitude for two Majorana neutrino production process in $e^+e^-$
 interaction is given by the contributions from six diagrams with gauge
  boson exchange in t,u and s channels (see Fig.1).
The contributions from Higgs exchange particles are negligible [10]
  and we do not consider them here.

  Full helicity amplitudes $M(\sigma \bar{\sigma};\lambda_1 \lambda_2)$
  for the process
  \begin{equation}
e^-(\sigma)+e^+(\bar{\sigma}) \rightarrow N_1(\lambda
  _1)+N_2(\lambda_2)
\end{equation}
are presented in Appendix of Refs[6] and [10].

  The CP effects are caused by phase factors that appear in the mixing
  matrices $K_{L,R}$ in t and u channels and $\Omega_{L,R}$ in s channel.
  To observe the influence of these phases two things must happen.
  First, different CP phases have to contribute to various Feynman diagrams
  from Fig.1, and second, the diagrams have to interfere so that at least
  two Feynman diagrams must contribute to the same helicity amplitude.
  The same mixing matrix elements give contributions to the $W_1$ ,
  $W_2$ exchange diagrams in t-u channels  $(K_{L,R})$ and $Z_1,Z_2$ bosons
exchange
  in s channel $(\Omega_{L,R})$. So even if these diagrams contribute to the
same helicity
   amlitude they do not interfere (of course there are also other suppression
   factors as the gauge boson mixing angles are small [6]).
   If the energy is large compared to the masses of neutrinos $N_1$ and $N_2$
then the t channel contributes to $M(- +;- +)$ (left-handed current) and $M(+
-;+ -)$
   (right-handed current) and the u channel gives contributions to $M(- +;+ -)$
    and $M(+ -;- +)$ amplitudes. We can see that at high energy there is no
interference
     between t and u channels [4]. The s-channel diagrams produce all four
     helicity amplitudes. So at high energy we can look for CP effects
resulting
from the interference between t-s and u-s channels.

  For the energy just above the production threshold there is no helicity
  suppression mechanism and final neutrinos with all helicity states can
   be produced by each channel diagram. These are the best conditions for
observing the
  CP violation effects.

  Another question is in what experimental observables the CP effects
  are visible. From the discussion presented above we can see that they can be
  looked for in polarized angular distribution. Unfortunately the cross
  sections, as we shall see, are too small to realize this possibility. And
what
  about the unpolarized angular distribution? If CP is conserved then the
  helicity amplitude satisfies the relation ($\Theta$ and $\phi$ are CM
  scattering angles)

\begin{eqnarray}
M(\sigma,\bar{\sigma};\lambda_1,\lambda_2;\Theta,\phi)&=&-\eta_{CP}^{\ast}(1)
\eta_{CP}^{\ast}(2)\times \nonumber \\
&&M(-\bar{\sigma},-\sigma;-\lambda_1,-\lambda_2;\pi-\Theta,\pi+\phi).
\end{eqnarray}
where $\eta_{CP}(i)$ are CP parities of the Majorana neutrinos. If we sum over
all
helicity the unpolarized angular distribution has forward-backward isotropy

\begin{equation}
\frac{d\sigma}{d\Omega}(\Theta,\phi)=\frac{d\sigma}{d\Omega}(\pi-\Theta,
\pi+\phi).
\end{equation}
Does it mean that anisotropy can be observed if CP is violated? Unfortunetely
not, at least
if we neglect the final state interaction. Without final state interaction from
CPT symmetry we can prove the relation
\begin{eqnarray}
M(\sigma,\bar{\sigma};\lambda_1,\lambda_2;\Theta,\phi)&=&-\eta_{CP}(1)
\eta_{CP}(2) e^{2i(\sigma-\bar{\sigma})(\pi+\phi)}\nonumber \\
&&M^{\ast}(-\bar{\sigma},-\sigma;-\lambda_1,-\lambda_2;\pi-\Theta,\pi+\phi)
\end{eqnarray}
from which the forward-backward isotropy also follows [11].
So the only observables where we can try to find the CP violation effect are
the total
cross sections. How big the effects can be?
There are six phases which cause the CP symmetry breaking. We do not try to
find the phase for which the effects of CP breaking is maximal. We take the
matrices $M_D$ and $M_R$ in the form
\begin{eqnarray*}
M_D&=&\left( \matrix{ 1. & 1. & .9 \cr
                      1. & 1. & .9 \cr
                      .9 & .9 & .95 } \right),
\end{eqnarray*}
and
\begin{eqnarray*}
M_R&=&\left( \matrix{ 150e^{i\alpha} & 10 & 20 \cr
                     10 & 200e^{i\beta} & 10 \cr
                     20 & 10 & 10^6e^{i\gamma} } \right),
\end{eqnarray*}
which produce a reasonable spectrum of light neutrinos. If we compare these
matrices with Eq.(12) we see that if only one or more phases ($\alpha,\beta$ or
$\gamma$) are not equal 0 or $\pi$ the CP is violated. Two heavy neutrinos with
masses $M_1\simeq 150 $ GeV and $M_2 \simeq 200$ GeV, almost independent
of the phases $\alpha,\beta$ and $\gamma$, result from our mass matrix.
We calculate the cross section for production of these neutrinos in $e^+e^-$
scattering
$$e^+e^- \rightarrow N_1(150)N_2(200). $$
The appropriate mixing matrix elements $(K_{L,R})_{1e},(K_{L,R})_{2e}$
and
$(\Omega_{L,R})_{12}$ depend on the phases $\alpha$ and $\beta$ and are
almost independent of the phase $\gamma$. For $\alpha=\beta=\gamma=0$
two neutrinos have equal CP parity and CP is conserved
\begin{equation}
\eta_{CP}(N_1)=\eta_{CP}(N_2)=+i.
\end{equation}
For $\alpha=\pi,\beta=\gamma=0$ CP is also conserved if we introduce the CP
parities
\begin{equation}
-\eta_{CP}(N_1)=\eta_{CP}(N_2)=+i.
\end{equation}
For any other values of phases CP is violated. The production cross sections
as energy functions are presented in Fig.2. Two factors affect the
behaviour of the cross section. First, there is real CP effect which causes
the different interference between various diagrams. Second, for different
phases different mixing matrix elements are obtained. In Fig.2 both these
effects are taken into account. To find the influence of CP interference only
we present in Fig.3 the cross sections for the same mixing matrix elements but
with all phases the same as in Fig.2. We can see that the influence of the CP
interference is very large. The cross section for production of two neutrinos
with opposite CP parity can be several times
bigger then the cross section for production of the same CP parity neutrinos.
The cross sections for the real CP breaking case are placed between two CP
conserving situations. We would like to stress that now the CP effect can be
quite large contrary to the Kobayashi-Maskawa mechanism in the quark sector.
In the lepton sector with Majorana neutrinos the changes in cross section which
result from CP breaking can be several times bigger than the cross section
itself. Unfortunately, the calculated cross sections are of the range of
several
femtobarns so the actual observation of the process for reasonable luminosity
will be difficult.
\section{Conclusions}
\ \ \
If Majorana neutrino are present in lepton sector the CP violation effect
can be very strong. For two heavy neutrinos production process $e^+e^-
\rightarrow N_1N_2$ the CP violation signals appear as an effect of t-u
channel interference just above the treshold and t-s, u-s channels interference
for higher energy. The angular distribution for unpolarized $e^+e^-$ beams
and without the measurement of the final neutrinos polarization has
forward-backward
symmetry even if CP is violated but the final state interaction may be
neglected.
The total cross section is the quantity which changes dramatically with various
CP violating parameters. Even if the change of total cross section is large
the cross section is small what makes the observation of this effect difficult.
\newline
$^{\dag}$e-mail gluza@usctoux1.cto.us.edu.pl  \\
$^{\ddagger}$e-mail zralek@usctoux1.cto.us.edu.pl
\section*{References}
\newcounter{bban}
\begin{list}
{$[{\ \arabic {bban}\ }]$}{\usecounter{bban}\setlength{\rightmargin}
{\leftmargin}}
\item M.Kobayashi and T.Maskawa,Prog.Theor.Phys.{\bf 49}(1973)652.
\item J.H.Christenson,J.W.Cronin,W.L.Fitch and R.Turlay, \newline
Phys.Rev.Lett.{\bf 13}(1964)138.
\item S.M.Bilenky,J.Hosek and S.Petcov,Phys.Letters 94{\bf B}(1980)495;
J.Schechter and L.Valle,Phys.Rev.{\bf D}22(1980)2227;\newline
M.Doi at al.,Phys.Letters 102{\bf B}(1981)323.
\item A.Barosso and J.Maalampi,Phys.Letters 132{\bf B}(1983)355;\newline
B.Kayser "CP effects when neutrinos are their own antiparticles" in "CP
violation" ed.by C.Jarlskog;World Scientific,Singapore (1989)p.334.
\item J.C.Pati and A.Salam,Phys.Rev.{\bf D}10,275(1974);\newline
R.N.Mohapatra and J.C.Pati,ibid.11,566(1975);11,2559(1975);
G.Senjanovic and R.N.Mohapatra,ibid.12,152(1975);\newline
G.Senjanovic,Nucl.Phys.{\bf B}153,334(1979).
\item N.G.Deshpande,J.F.Gunion,B.Kayser and F.Olness,\newline
Phys.Rev.{\bf D}44(1991)837;\newline
J.Gluza and M.Zra\l ek,Phys.Rev.{\bf D}48(1993)5093.
\item G.C.Branco and M.N.Rabelo,Phys.Letters{\bf B}173(1986)313;
J.Bernabeu,G.C.Branco and M.Gronau,Phys.Lett.{\bf B}169(1986)243.
\item For the review see: S.M.Bilenky and S.T.Petcov,Review of Mod.Phys.59
(1987)671.
\item J.Bernabeu and P.Pascual,Nucl.Phys.{\bf B}228(1983)21;\newline
B.Kayser,Phys.Rev.{\bf D}30(1984)1023.
\item J.Gluza and M.Zra\l ek;Silesian Univ.Preprint,submitted to ICHEP
Ref.0123,
June 1994.
\item S.Petcov Phys.Lett.178{\bf B}(1986)57.
\end{list}
\section*{Figure Captions}
\newcounter{bean}
\begin{list}
{\bf Fig.\arabic {bban}}{\usecounter{bban}\setlength{\rightmargin}
{\leftmargin}}
\item Diagrams with gauge boson exchange which describe the process $e^-e^+
\rightarrow N_1N_2$
in the left-right symmetric model on the tree level.
\item CP and mixing matrix effects for the $e^-e^+ \rightarrow
N_1(150)N_2(200)$
production. Solid line is for $\alpha=\beta=\gamma=0$,
dotted line is for $\alpha=2.0,\beta=\gamma=0$ and the third line (solid with
asterisks) is
for $\alpha=\pi,\beta=\gamma=0$ phases. The other L-R model parameters which we
used are the following:$M_{W_2}=1500$ GeV,$\beta=\frac{M_{W_1}^2}{M_{W_2}^2}$,
$M_{Z_2}^2=\frac{2cos^4{\Theta_W}M_{Z_1}^2}{\cos{2\Theta_W}\beta}$ ,$\;\xi
=\beta$,
$\phi=-\frac{{(\cos{2\Theta_W})}^{3/2}}{2\cos^4{\Theta_W}}\beta$ (see Ref.[6]).
\item The effect of CP violation only on the $e^-e^+ \rightarrow
N_1(150)N_2(200)$ production.
Absolute values of mixing matrix elements are the same as the ones for solid
line
in Fig.2 $({(K_L)}_{1e}=.00535,{(K_R)}_{1e}=.9819,{(K_L)}_{2e}=.0058,
{(K_R)}_{2e}=.189,{(\Omega_L)}_{12}=-{(\Omega_R)}_{12}=.00009 )$.
Solid (dotted) line is for opposite (the same)
CP parity of neutrinos (Eqs.27 and 26). Solid line with asterisks is for
$\alpha=2.0,
\beta=\gamma=0$, the same as in Fig.2.
\end{list}

\newpage
\input FEYNMAN
\bigphotons
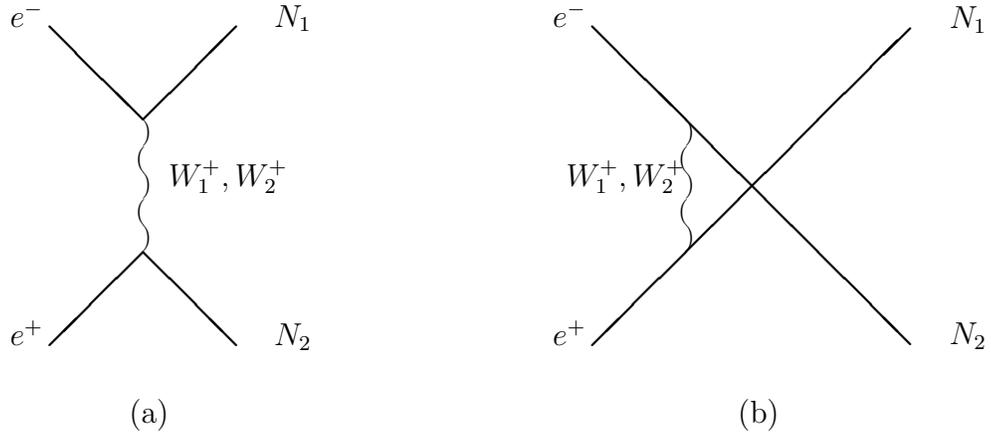
\centerline{Fig.1. Feynman diagrams with gauge boson particles}
\begin{picture}(20000,20000)(0,0)
\thicklines
\drawline\fermion[\SE\REG](1500,15000)[5000]
\put(0,\pfronty){$e^-$}
\drawline\photon[\S\REG](\pbackx,\pbacky)[5]
\put(6000,\pmidy){$W_1^+,W_2^+$}
\drawline\fermion[\SW\REG](\photonbackx,\photonbacky)[5000]
\put(0,\pbacky){$e^+$}
\drawline\fermion[\SE\REG](\photonbackx,\photonbacky)[5000]
\put(10000,\pbacky){$N_2$}
\drawline\fermion[\NE\REG](\photonfrontx,\photonfronty)[5000]
\put(10000,\pbacky){$N_1$}
\put(4500,0){(a)}
\end{picture}
\hskip .1cm
\begin{picture}(20000,20000)(0,0)
\thicklines
\drawline\fermion[\SE\REG](1500,15000)[5000]
\put(0,\pfronty){$e^-$}
\global\seglength=1000
\global\gaplength=200
\drawline\photon[\S\REG](\pbackx,\pbacky)[5]
\put(500,\pmidy){$W_1^+,W_2^+$}
\drawline\fermion[\SW\REG](\photonbackx,\photonbacky)[5000]
\put(0,\pbacky){$e^+$}
\drawline\fermion[\NE\REG](\photonbackx,\photonbacky)[12000]
\put(15000,\pbacky){$N_1$}
\drawline\fermion[\SE\REG](\photonfrontx,\photonfronty)[12000]
\put(15000,\pbacky){$N_2$}
\put(7000,0){(b)}
\end{picture}
\begin{picture}(20000,20000)(-8000,-8000)
\thicklines
\drawline\photon[\E\REG](5000,4000)[9]
\put(\pfrontx,5000){$\;\;\;\;\;Z_1^0,Z_2^0$}
\drawline\fermion[\SW\REG](\photonfrontx,\photonfronty)[5000]
\put(0,\pbacky){$e^+$}
\drawline\fermion[\NW\REG](\photonfrontx,\photonfronty)[5000]
\put(0,\pbacky){$e^-$}
\drawline\fermion[\SE\REG](\photonbackx,\photonbacky)[5000]
\put(19000,\pbacky){$N_2$}
\drawline\fermion[\NE\REG](\photonbackx,\photonbacky)[5000]
\put(19000,\pbacky){$N_1$}
\put(9000,-3000){(c)}
\end{picture}
\newpage
\setlength{\unitlength}{0.240900pt}
\ifx\plotpoint\undefined\newsavebox{\plotpoint}\fi
\sbox{\plotpoint}{\rule[-0.200pt]{0.400pt}{0.400pt}}%
\begin{center}
\begin{picture}(1500,900)(0,0)
\font\gnuplot=cmr10 at 10pt
\gnuplot
\sbox{\plotpoint}{\rule[-0.200pt]{0.400pt}{0.400pt}}%
\put(220.0,113.0){\rule[-0.200pt]{292.934pt}{0.400pt}}
\put(220.0,113.0){\rule[-0.200pt]{0.400pt}{184.048pt}}
\put(220.0,113.0){\rule[-0.200pt]{4.818pt}{0.400pt}}
\put(198,113){\makebox(0,0)[r]{0}}
\put(1416.0,113.0){\rule[-0.200pt]{4.818pt}{0.400pt}}
\put(220.0,209.0){\rule[-0.200pt]{4.818pt}{0.400pt}}
\put(198,209){\makebox(0,0)[r]{0.5}}
\put(1416.0,209.0){\rule[-0.200pt]{4.818pt}{0.400pt}}
\put(220.0,304.0){\rule[-0.200pt]{4.818pt}{0.400pt}}
\put(198,304){\makebox(0,0)[r]{1}}
\put(1416.0,304.0){\rule[-0.200pt]{4.818pt}{0.400pt}}
\put(220.0,400.0){\rule[-0.200pt]{4.818pt}{0.400pt}}
\put(198,400){\makebox(0,0)[r]{1.5}}
\put(1416.0,400.0){\rule[-0.200pt]{4.818pt}{0.400pt}}
\put(220.0,495.0){\rule[-0.200pt]{4.818pt}{0.400pt}}
\put(198,495){\makebox(0,0)[r]{2}}
\put(1416.0,495.0){\rule[-0.200pt]{4.818pt}{0.400pt}}
\put(220.0,591.0){\rule[-0.200pt]{4.818pt}{0.400pt}}
\put(198,591){\makebox(0,0)[r]{2.5}}
\put(1416.0,591.0){\rule[-0.200pt]{4.818pt}{0.400pt}}
\put(220.0,686.0){\rule[-0.200pt]{4.818pt}{0.400pt}}
\put(198,686){\makebox(0,0)[r]{3}}
\put(1416.0,686.0){\rule[-0.200pt]{4.818pt}{0.400pt}}
\put(220.0,782.0){\rule[-0.200pt]{4.818pt}{0.400pt}}
\put(198,782){\makebox(0,0)[r]{3.5}}
\put(1416.0,782.0){\rule[-0.200pt]{4.818pt}{0.400pt}}
\put(220.0,877.0){\rule[-0.200pt]{4.818pt}{0.400pt}}
\put(198,877){\makebox(0,0)[r]{4}}
\put(1416.0,877.0){\rule[-0.200pt]{4.818pt}{0.400pt}}
\put(220.0,113.0){\rule[-0.200pt]{0.400pt}{4.818pt}}
\put(220,68){\makebox(0,0){0}}
\put(220.0,857.0){\rule[-0.200pt]{0.400pt}{4.818pt}}
\put(342.0,113.0){\rule[-0.200pt]{0.400pt}{4.818pt}}
\put(342,68){\makebox(0,0){500}}
\put(342.0,857.0){\rule[-0.200pt]{0.400pt}{4.818pt}}
\put(463.0,113.0){\rule[-0.200pt]{0.400pt}{4.818pt}}
\put(463,68){\makebox(0,0){1000}}
\put(463.0,857.0){\rule[-0.200pt]{0.400pt}{4.818pt}}
\put(585.0,113.0){\rule[-0.200pt]{0.400pt}{4.818pt}}
\put(585,68){\makebox(0,0){1500}}
\put(585.0,857.0){\rule[-0.200pt]{0.400pt}{4.818pt}}
\put(706.0,113.0){\rule[-0.200pt]{0.400pt}{4.818pt}}
\put(706,68){\makebox(0,0){2000}}
\put(706.0,857.0){\rule[-0.200pt]{0.400pt}{4.818pt}}
\put(828.0,113.0){\rule[-0.200pt]{0.400pt}{4.818pt}}
\put(828,68){\makebox(0,0){2500}}
\put(828.0,857.0){\rule[-0.200pt]{0.400pt}{4.818pt}}
\put(950.0,113.0){\rule[-0.200pt]{0.400pt}{4.818pt}}
\put(950,68){\makebox(0,0){3000}}
\put(950.0,857.0){\rule[-0.200pt]{0.400pt}{4.818pt}}
\put(1071.0,113.0){\rule[-0.200pt]{0.400pt}{4.818pt}}
\put(1071,68){\makebox(0,0){3500}}
\put(1071.0,857.0){\rule[-0.200pt]{0.400pt}{4.818pt}}
\put(1193.0,113.0){\rule[-0.200pt]{0.400pt}{4.818pt}}
\put(1193,68){\makebox(0,0){4000}}
\put(1193.0,857.0){\rule[-0.200pt]{0.400pt}{4.818pt}}
\put(1314.0,113.0){\rule[-0.200pt]{0.400pt}{4.818pt}}
\put(1314,68){\makebox(0,0){4500}}
\put(1314.0,857.0){\rule[-0.200pt]{0.400pt}{4.818pt}}
\put(1436.0,113.0){\rule[-0.200pt]{0.400pt}{4.818pt}}
\put(1436,68){\makebox(0,0){5000}}
\put(1436.0,857.0){\rule[-0.200pt]{0.400pt}{4.818pt}}
\put(220.0,113.0){\rule[-0.200pt]{292.934pt}{0.400pt}}
\put(1436.0,113.0){\rule[-0.200pt]{0.400pt}{184.048pt}}
\put(220.0,877.0){\rule[-0.200pt]{292.934pt}{0.400pt}}
\put(45,495){\makebox(0,0){$\sigma$[fb]}}
\put(828,950){\makebox(0,0){Fig.2}}
\put(828,-50){\makebox(0,0){$\sqrt{s}$[GeV]}}
\put(220.0,113.0){\rule[-0.200pt]{0.400pt}{184.048pt}}
\put(305,113){\usebox{\plotpoint}}
\multiput(305.58,113.00)(0.498,0.594){103}{\rule{0.120pt}{0.575pt}}
\multiput(304.17,113.00)(53.000,61.806){2}{\rule{0.400pt}{0.288pt}}
\multiput(358.58,176.00)(0.498,0.718){103}{\rule{0.120pt}{0.674pt}}
\multiput(357.17,176.00)(53.000,74.602){2}{\rule{0.400pt}{0.337pt}}
\multiput(411.58,252.00)(0.498,0.519){101}{\rule{0.120pt}{0.515pt}}
\multiput(410.17,252.00)(52.000,52.930){2}{\rule{0.400pt}{0.258pt}}
\multiput(463.00,306.58)(0.918,0.497){55}{\rule{0.831pt}{0.120pt}}
\multiput(463.00,305.17)(51.275,29.000){2}{\rule{0.416pt}{0.400pt}}
\multiput(516.00,335.58)(2.737,0.491){17}{\rule{2.220pt}{0.118pt}}
\multiput(516.00,334.17)(48.392,10.000){2}{\rule{1.110pt}{0.400pt}}
\put(569,344.67){\rule{12.768pt}{0.400pt}}
\multiput(569.00,344.17)(26.500,1.000){2}{\rule{6.384pt}{0.400pt}}
\put(622,345.67){\rule{12.768pt}{0.400pt}}
\multiput(622.00,345.17)(26.500,1.000){2}{\rule{6.384pt}{0.400pt}}
\multiput(675.00,347.59)(5.719,0.477){7}{\rule{4.260pt}{0.115pt}}
\multiput(675.00,346.17)(43.158,5.000){2}{\rule{2.130pt}{0.400pt}}
\multiput(727.00,352.58)(2.263,0.492){21}{\rule{1.867pt}{0.119pt}}
\multiput(727.00,351.17)(49.126,12.000){2}{\rule{0.933pt}{0.400pt}}
\multiput(780.00,364.58)(1.338,0.496){37}{\rule{1.160pt}{0.119pt}}
\multiput(780.00,363.17)(50.592,20.000){2}{\rule{0.580pt}{0.400pt}}
\multiput(833.00,384.58)(0.987,0.497){51}{\rule{0.885pt}{0.120pt}}
\multiput(833.00,383.17)(51.163,27.000){2}{\rule{0.443pt}{0.400pt}}
\multiput(886.00,411.58)(0.831,0.497){61}{\rule{0.762pt}{0.120pt}}
\multiput(886.00,410.17)(51.417,32.000){2}{\rule{0.381pt}{0.400pt}}
\multiput(939.00,443.58)(0.704,0.498){71}{\rule{0.662pt}{0.120pt}}
\multiput(939.00,442.17)(50.626,37.000){2}{\rule{0.331pt}{0.400pt}}
\multiput(991.00,480.58)(0.647,0.498){79}{\rule{0.617pt}{0.120pt}}
\multiput(991.00,479.17)(51.719,41.000){2}{\rule{0.309pt}{0.400pt}}
\multiput(1044.00,521.58)(0.616,0.498){83}{\rule{0.593pt}{0.120pt}}
\multiput(1044.00,520.17)(51.769,43.000){2}{\rule{0.297pt}{0.400pt}}
\multiput(1097.00,564.58)(0.602,0.498){85}{\rule{0.582pt}{0.120pt}}
\multiput(1097.00,563.17)(51.792,44.000){2}{\rule{0.291pt}{0.400pt}}
\multiput(1150.00,608.58)(0.589,0.498){87}{\rule{0.571pt}{0.120pt}}
\multiput(1150.00,607.17)(51.815,45.000){2}{\rule{0.286pt}{0.400pt}}
\multiput(1203.00,653.58)(0.565,0.498){89}{\rule{0.552pt}{0.120pt}}
\multiput(1203.00,652.17)(50.854,46.000){2}{\rule{0.276pt}{0.400pt}}
\multiput(1255.00,699.58)(0.589,0.498){87}{\rule{0.571pt}{0.120pt}}
\multiput(1255.00,698.17)(51.815,45.000){2}{\rule{0.286pt}{0.400pt}}
\multiput(1308.00,744.58)(0.602,0.498){85}{\rule{0.582pt}{0.120pt}}
\multiput(1308.00,743.17)(51.792,44.000){2}{\rule{0.291pt}{0.400pt}}
\put(305,113){\rule{1pt}{1pt}}
\put(358,147){\rule{1pt}{1pt}}
\put(411,182){\rule{1pt}{1pt}}
\put(463,210){\rule{1pt}{1pt}}
\put(516,228){\rule{1pt}{1pt}}
\put(569,237){\rule{1pt}{1pt}}
\put(622,242){\rule{1pt}{1pt}}
\put(675,246){\rule{1pt}{1pt}}
\put(727,249){\rule{1pt}{1pt}}
\put(780,255){\rule{1pt}{1pt}}
\put(833,262){\rule{1pt}{1pt}}
\put(886,271){\rule{1pt}{1pt}}
\put(939,281){\rule{1pt}{1pt}}
\put(991,293){\rule{1pt}{1pt}}
\put(1044,306){\rule{1pt}{1pt}}
\put(1097,320){\rule{1pt}{1pt}}
\put(1150,334){\rule{1pt}{1pt}}
\put(1203,349){\rule{1pt}{1pt}}
\put(1255,363){\rule{1pt}{1pt}}
\put(1308,377){\rule{1pt}{1pt}}
\put(1361,391){\rule{1pt}{1pt}}
\sbox{\plotpoint}{\rule[-0.400pt]{0.800pt}{0.800pt}}%
\put(305,113){\usebox{\plotpoint}}
\multiput(305.00,114.41)(1.288,0.505){35}{\rule{2.219pt}{0.122pt}}
\multiput(305.00,111.34)(48.394,21.000){2}{\rule{1.110pt}{0.800pt}}
\multiput(358.00,135.41)(1.838,0.508){23}{\rule{3.027pt}{0.122pt}}
\multiput(358.00,132.34)(46.718,15.000){2}{\rule{1.513pt}{0.800pt}}
\multiput(411.00,150.41)(1.803,0.508){23}{\rule{2.973pt}{0.122pt}}
\multiput(411.00,147.34)(45.829,15.000){2}{\rule{1.487pt}{0.800pt}}
\multiput(463.00,165.41)(2.146,0.509){19}{\rule{3.462pt}{0.123pt}}
\multiput(463.00,162.34)(45.815,13.000){2}{\rule{1.731pt}{0.800pt}}
\multiput(516.00,178.40)(2.881,0.514){13}{\rule{4.440pt}{0.124pt}}
\multiput(516.00,175.34)(43.785,10.000){2}{\rule{2.220pt}{0.800pt}}
\multiput(569.00,188.40)(4.504,0.526){7}{\rule{6.257pt}{0.127pt}}
\multiput(569.00,185.34)(40.013,7.000){2}{\rule{3.129pt}{0.800pt}}
\multiput(622.00,195.38)(8.484,0.560){3}{\rule{8.680pt}{0.135pt}}
\multiput(622.00,192.34)(34.984,5.000){2}{\rule{4.340pt}{0.800pt}}
\put(675,198.84){\rule{12.527pt}{0.800pt}}
\multiput(675.00,197.34)(26.000,3.000){2}{\rule{6.263pt}{0.800pt}}
\put(727,201.34){\rule{12.768pt}{0.800pt}}
\multiput(727.00,200.34)(26.500,2.000){2}{\rule{6.384pt}{0.800pt}}
\put(780,202.84){\rule{12.768pt}{0.800pt}}
\multiput(780.00,202.34)(26.500,1.000){2}{\rule{6.384pt}{0.800pt}}
\put(886,202.84){\rule{12.768pt}{0.800pt}}
\multiput(886.00,203.34)(26.500,-1.000){2}{\rule{6.384pt}{0.800pt}}
\put(939,201.84){\rule{12.527pt}{0.800pt}}
\multiput(939.00,202.34)(26.000,-1.000){2}{\rule{6.263pt}{0.800pt}}
\put(991,200.84){\rule{12.768pt}{0.800pt}}
\multiput(991.00,201.34)(26.500,-1.000){2}{\rule{6.384pt}{0.800pt}}
\put(1044,199.84){\rule{12.768pt}{0.800pt}}
\multiput(1044.00,200.34)(26.500,-1.000){2}{\rule{6.384pt}{0.800pt}}
\put(1097,198.34){\rule{12.768pt}{0.800pt}}
\multiput(1097.00,199.34)(26.500,-2.000){2}{\rule{6.384pt}{0.800pt}}
\put(1150,196.84){\rule{12.768pt}{0.800pt}}
\multiput(1150.00,197.34)(26.500,-1.000){2}{\rule{6.384pt}{0.800pt}}
\put(1203,195.34){\rule{12.527pt}{0.800pt}}
\multiput(1203.00,196.34)(26.000,-2.000){2}{\rule{6.263pt}{0.800pt}}
\put(1255,193.84){\rule{12.768pt}{0.800pt}}
\multiput(1255.00,194.34)(26.500,-1.000){2}{\rule{6.384pt}{0.800pt}}
\put(1308,192.34){\rule{12.768pt}{0.800pt}}
\multiput(1308.00,193.34)(26.500,-2.000){2}{\rule{6.384pt}{0.800pt}}
\put(305,113){\raisebox{-.8pt}{\makebox(0,0){$\ast$}}}
\put(358,134){\raisebox{-.8pt}{\makebox(0,0){$\ast$}}}
\put(411,149){\raisebox{-.8pt}{\makebox(0,0){$\ast$}}}
\put(463,164){\raisebox{-.8pt}{\makebox(0,0){$\ast$}}}
\put(516,177){\raisebox{-.8pt}{\makebox(0,0){$\ast$}}}
\put(569,187){\raisebox{-.8pt}{\makebox(0,0){$\ast$}}}
\put(622,194){\raisebox{-.8pt}{\makebox(0,0){$\ast$}}}
\put(675,199){\raisebox{-.8pt}{\makebox(0,0){$\ast$}}}
\put(727,202){\raisebox{-.8pt}{\makebox(0,0){$\ast$}}}
\put(780,204){\raisebox{-.8pt}{\makebox(0,0){$\ast$}}}
\put(833,205){\raisebox{-.8pt}{\makebox(0,0){$\ast$}}}
\put(886,205){\raisebox{-.8pt}{\makebox(0,0){$\ast$}}}
\put(939,204){\raisebox{-.8pt}{\makebox(0,0){$\ast$}}}
\put(991,203){\raisebox{-.8pt}{\makebox(0,0){$\ast$}}}
\put(1044,202){\raisebox{-.8pt}{\makebox(0,0){$\ast$}}}
\put(1097,201){\raisebox{-.8pt}{\makebox(0,0){$\ast$}}}
\put(1150,199){\raisebox{-.8pt}{\makebox(0,0){$\ast$}}}
\put(1203,198){\raisebox{-.8pt}{\makebox(0,0){$\ast$}}}
\put(1255,196){\raisebox{-.8pt}{\makebox(0,0){$\ast$}}}
\put(1308,195){\raisebox{-.8pt}{\makebox(0,0){$\ast$}}}
\put(1361,193){\raisebox{-.8pt}{\makebox(0,0){$\ast$}}}
\put(833.0,205.0){\rule[-0.400pt]{12.768pt}{0.800pt}}
\end{picture}
\end{center}
\newpage
\setlength{\unitlength}{0.240900pt}
\ifx\plotpoint\undefined\newsavebox{\plotpoint}\fi
\sbox{\plotpoint}{\rule[-0.200pt]{0.400pt}{0.400pt}}%
\begin{center}
\begin{picture}(1500,900)(0,0)
\font\gnuplot=cmr10 at 10pt
\gnuplot
\sbox{\plotpoint}{\rule[-0.200pt]{0.400pt}{0.400pt}}%
\put(220.0,113.0){\rule[-0.200pt]{292.934pt}{0.400pt}}
\put(220.0,113.0){\rule[-0.200pt]{0.400pt}{184.048pt}}
\put(220.0,113.0){\rule[-0.200pt]{4.818pt}{0.400pt}}
\put(198,113){\makebox(0,0)[r]{0}}
\put(1416.0,113.0){\rule[-0.200pt]{4.818pt}{0.400pt}}
\put(220.0,266.0){\rule[-0.200pt]{4.818pt}{0.400pt}}
\put(198,266){\makebox(0,0)[r]{5}}
\put(1416.0,266.0){\rule[-0.200pt]{4.818pt}{0.400pt}}
\put(220.0,419.0){\rule[-0.200pt]{4.818pt}{0.400pt}}
\put(198,419){\makebox(0,0)[r]{10}}
\put(1416.0,419.0){\rule[-0.200pt]{4.818pt}{0.400pt}}
\put(220.0,571.0){\rule[-0.200pt]{4.818pt}{0.400pt}}
\put(198,571){\makebox(0,0)[r]{15}}
\put(1416.0,571.0){\rule[-0.200pt]{4.818pt}{0.400pt}}
\put(220.0,724.0){\rule[-0.200pt]{4.818pt}{0.400pt}}
\put(198,724){\makebox(0,0)[r]{20}}
\put(1416.0,724.0){\rule[-0.200pt]{4.818pt}{0.400pt}}
\put(220.0,877.0){\rule[-0.200pt]{4.818pt}{0.400pt}}
\put(198,877){\makebox(0,0)[r]{25}}
\put(1416.0,877.0){\rule[-0.200pt]{4.818pt}{0.400pt}}
\put(220.0,113.0){\rule[-0.200pt]{0.400pt}{4.818pt}}
\put(220,68){\makebox(0,0){0}}
\put(220.0,857.0){\rule[-0.200pt]{0.400pt}{4.818pt}}
\put(342.0,113.0){\rule[-0.200pt]{0.400pt}{4.818pt}}
\put(342,68){\makebox(0,0){500}}
\put(342.0,857.0){\rule[-0.200pt]{0.400pt}{4.818pt}}
\put(463.0,113.0){\rule[-0.200pt]{0.400pt}{4.818pt}}
\put(463,68){\makebox(0,0){1000}}
\put(463.0,857.0){\rule[-0.200pt]{0.400pt}{4.818pt}}
\put(585.0,113.0){\rule[-0.200pt]{0.400pt}{4.818pt}}
\put(585,68){\makebox(0,0){1500}}
\put(585.0,857.0){\rule[-0.200pt]{0.400pt}{4.818pt}}
\put(706.0,113.0){\rule[-0.200pt]{0.400pt}{4.818pt}}
\put(706,68){\makebox(0,0){2000}}
\put(706.0,857.0){\rule[-0.200pt]{0.400pt}{4.818pt}}
\put(828.0,113.0){\rule[-0.200pt]{0.400pt}{4.818pt}}
\put(828,68){\makebox(0,0){2500}}
\put(828.0,857.0){\rule[-0.200pt]{0.400pt}{4.818pt}}
\put(950.0,113.0){\rule[-0.200pt]{0.400pt}{4.818pt}}
\put(950,68){\makebox(0,0){3000}}
\put(950.0,857.0){\rule[-0.200pt]{0.400pt}{4.818pt}}
\put(1071.0,113.0){\rule[-0.200pt]{0.400pt}{4.818pt}}
\put(1071,68){\makebox(0,0){3500}}
\put(1071.0,857.0){\rule[-0.200pt]{0.400pt}{4.818pt}}
\put(1193.0,113.0){\rule[-0.200pt]{0.400pt}{4.818pt}}
\put(1193,68){\makebox(0,0){4000}}
\put(1193.0,857.0){\rule[-0.200pt]{0.400pt}{4.818pt}}
\put(1314.0,113.0){\rule[-0.200pt]{0.400pt}{4.818pt}}
\put(1314,68){\makebox(0,0){4500}}
\put(1314.0,857.0){\rule[-0.200pt]{0.400pt}{4.818pt}}
\put(1436.0,113.0){\rule[-0.200pt]{0.400pt}{4.818pt}}
\put(1436,68){\makebox(0,0){5000}}
\put(1436.0,857.0){\rule[-0.200pt]{0.400pt}{4.818pt}}
\put(220.0,113.0){\rule[-0.200pt]{292.934pt}{0.400pt}}
\put(1436.0,113.0){\rule[-0.200pt]{0.400pt}{184.048pt}}
\put(220.0,877.0){\rule[-0.200pt]{292.934pt}{0.400pt}}
\put(45,495){\makebox(0,0){$\sigma$[fb]}}
\put(828,950){\makebox(0,0){Fig.3}}
\put(828,-50){\makebox(0,0){$\sqrt{s}$[GeV]}}
\put(220.0,113.0){\rule[-0.200pt]{0.400pt}{184.048pt}}
\put(305,113){\usebox{\plotpoint}}
\multiput(305.58,113.00)(0.498,1.335){103}{\rule{0.120pt}{1.164pt}}
\multiput(304.17,113.00)(53.000,138.584){2}{\rule{0.400pt}{0.582pt}}
\multiput(358.58,254.00)(0.498,1.022){103}{\rule{0.120pt}{0.915pt}}
\multiput(357.17,254.00)(53.000,106.101){2}{\rule{0.400pt}{0.458pt}}
\multiput(411.58,362.00)(0.498,0.974){101}{\rule{0.120pt}{0.877pt}}
\multiput(410.17,362.00)(52.000,99.180){2}{\rule{0.400pt}{0.438pt}}
\multiput(463.58,463.00)(0.498,0.813){103}{\rule{0.120pt}{0.749pt}}
\multiput(462.17,463.00)(53.000,84.445){2}{\rule{0.400pt}{0.375pt}}
\multiput(516.58,549.00)(0.498,0.651){103}{\rule{0.120pt}{0.621pt}}
\multiput(515.17,549.00)(53.000,67.712){2}{\rule{0.400pt}{0.310pt}}
\multiput(569.00,618.58)(0.529,0.498){97}{\rule{0.524pt}{0.120pt}}
\multiput(569.00,617.17)(51.912,50.000){2}{\rule{0.262pt}{0.400pt}}
\multiput(622.00,668.58)(0.759,0.498){67}{\rule{0.706pt}{0.120pt}}
\multiput(622.00,667.17)(51.535,35.000){2}{\rule{0.353pt}{0.400pt}}
\multiput(675.00,703.58)(1.192,0.496){41}{\rule{1.045pt}{0.120pt}}
\multiput(675.00,702.17)(49.830,22.000){2}{\rule{0.523pt}{0.400pt}}
\multiput(727.00,725.58)(2.263,0.492){21}{\rule{1.867pt}{0.119pt}}
\multiput(727.00,724.17)(49.126,12.000){2}{\rule{0.933pt}{0.400pt}}
\multiput(780.00,737.59)(5.831,0.477){7}{\rule{4.340pt}{0.115pt}}
\multiput(780.00,736.17)(43.992,5.000){2}{\rule{2.170pt}{0.400pt}}
\multiput(886.00,740.94)(7.646,-0.468){5}{\rule{5.400pt}{0.113pt}}
\multiput(886.00,741.17)(41.792,-4.000){2}{\rule{2.700pt}{0.400pt}}
\multiput(939.00,736.93)(3.925,-0.485){11}{\rule{3.071pt}{0.117pt}}
\multiput(939.00,737.17)(45.625,-7.000){2}{\rule{1.536pt}{0.400pt}}
\multiput(991.00,729.93)(3.465,-0.488){13}{\rule{2.750pt}{0.117pt}}
\multiput(991.00,730.17)(47.292,-8.000){2}{\rule{1.375pt}{0.400pt}}
\multiput(1044.00,721.92)(2.737,-0.491){17}{\rule{2.220pt}{0.118pt}}
\multiput(1044.00,722.17)(48.392,-10.000){2}{\rule{1.110pt}{0.400pt}}
\multiput(1097.00,711.92)(2.737,-0.491){17}{\rule{2.220pt}{0.118pt}}
\multiput(1097.00,712.17)(48.392,-10.000){2}{\rule{1.110pt}{0.400pt}}
\multiput(1150.00,701.92)(2.737,-0.491){17}{\rule{2.220pt}{0.118pt}}
\multiput(1150.00,702.17)(48.392,-10.000){2}{\rule{1.110pt}{0.400pt}}
\multiput(1203.00,691.92)(2.684,-0.491){17}{\rule{2.180pt}{0.118pt}}
\multiput(1203.00,692.17)(47.475,-10.000){2}{\rule{1.090pt}{0.400pt}}
\multiput(1255.00,681.92)(2.737,-0.491){17}{\rule{2.220pt}{0.118pt}}
\multiput(1255.00,682.17)(48.392,-10.000){2}{\rule{1.110pt}{0.400pt}}
\multiput(1308.00,671.92)(2.737,-0.491){17}{\rule{2.220pt}{0.118pt}}
\multiput(1308.00,672.17)(48.392,-10.000){2}{\rule{1.110pt}{0.400pt}}
\put(833.0,742.0){\rule[-0.200pt]{12.768pt}{0.400pt}}
\put(305,113){\rule{1pt}{1pt}}
\put(358,123){\rule{1pt}{1pt}}
\put(411,135){\rule{1pt}{1pt}}
\put(463,144){\rule{1pt}{1pt}}
\put(516,148){\rule{1pt}{1pt}}
\put(569,150){\rule{1pt}{1pt}}
\put(622,150){\rule{1pt}{1pt}}
\put(675,150){\rule{1pt}{1pt}}
\put(727,151){\rule{1pt}{1pt}}
\put(780,153){\rule{1pt}{1pt}}
\put(833,156){\rule{1pt}{1pt}}
\put(886,161){\rule{1pt}{1pt}}
\put(939,166){\rule{1pt}{1pt}}
\put(991,172){\rule{1pt}{1pt}}
\put(1044,178){\rule{1pt}{1pt}}
\put(1097,185){\rule{1pt}{1pt}}
\put(1150,192){\rule{1pt}{1pt}}
\put(1203,199){\rule{1pt}{1pt}}
\put(1255,207){\rule{1pt}{1pt}}
\put(1308,214){\rule{1pt}{1pt}}
\put(1361,221){\rule{1pt}{1pt}}
\sbox{\plotpoint}{\rule[-0.400pt]{0.800pt}{0.800pt}}%
\put(305,113){\usebox{\plotpoint}}
\multiput(305.00,114.41)(1.715,0.507){25}{\rule{2.850pt}{0.122pt}}
\multiput(305.00,111.34)(47.085,16.000){2}{\rule{1.425pt}{0.800pt}}
\multiput(358.00,130.41)(1.838,0.508){23}{\rule{3.027pt}{0.122pt}}
\multiput(358.00,127.34)(46.718,15.000){2}{\rule{1.513pt}{0.800pt}}
\multiput(411.00,145.41)(2.105,0.509){19}{\rule{3.400pt}{0.123pt}}
\multiput(411.00,142.34)(44.943,13.000){2}{\rule{1.700pt}{0.800pt}}
\multiput(463.00,158.40)(3.771,0.520){9}{\rule{5.500pt}{0.125pt}}
\multiput(463.00,155.34)(41.584,8.000){2}{\rule{2.750pt}{0.800pt}}
\put(516,165.34){\rule{10.800pt}{0.800pt}}
\multiput(516.00,163.34)(30.584,4.000){2}{\rule{5.400pt}{0.800pt}}
\put(569,168.84){\rule{12.768pt}{0.800pt}}
\multiput(569.00,167.34)(26.500,3.000){2}{\rule{6.384pt}{0.800pt}}
\put(622,170.84){\rule{12.768pt}{0.800pt}}
\multiput(622.00,170.34)(26.500,1.000){2}{\rule{6.384pt}{0.800pt}}
\put(675,172.34){\rule{12.527pt}{0.800pt}}
\multiput(675.00,171.34)(26.000,2.000){2}{\rule{6.263pt}{0.800pt}}
\put(727,174.34){\rule{12.768pt}{0.800pt}}
\multiput(727.00,173.34)(26.500,2.000){2}{\rule{6.384pt}{0.800pt}}
\put(780,176.84){\rule{12.768pt}{0.800pt}}
\multiput(780.00,175.34)(26.500,3.000){2}{\rule{6.384pt}{0.800pt}}
\put(833,180.34){\rule{10.800pt}{0.800pt}}
\multiput(833.00,178.34)(30.584,4.000){2}{\rule{5.400pt}{0.800pt}}
\multiput(886.00,185.38)(8.484,0.560){3}{\rule{8.680pt}{0.135pt}}
\multiput(886.00,182.34)(34.984,5.000){2}{\rule{4.340pt}{0.800pt}}
\multiput(939.00,190.39)(5.597,0.536){5}{\rule{7.133pt}{0.129pt}}
\multiput(939.00,187.34)(37.194,6.000){2}{\rule{3.567pt}{0.800pt}}
\multiput(991.00,196.39)(5.709,0.536){5}{\rule{7.267pt}{0.129pt}}
\multiput(991.00,193.34)(37.918,6.000){2}{\rule{3.633pt}{0.800pt}}
\multiput(1044.00,202.39)(5.709,0.536){5}{\rule{7.267pt}{0.129pt}}
\multiput(1044.00,199.34)(37.918,6.000){2}{\rule{3.633pt}{0.800pt}}
\multiput(1097.00,208.39)(5.709,0.536){5}{\rule{7.267pt}{0.129pt}}
\multiput(1097.00,205.34)(37.918,6.000){2}{\rule{3.633pt}{0.800pt}}
\multiput(1150.00,214.40)(4.504,0.526){7}{\rule{6.257pt}{0.127pt}}
\multiput(1150.00,211.34)(40.013,7.000){2}{\rule{3.129pt}{0.800pt}}
\multiput(1203.00,221.39)(5.597,0.536){5}{\rule{7.133pt}{0.129pt}}
\multiput(1203.00,218.34)(37.194,6.000){2}{\rule{3.567pt}{0.800pt}}
\multiput(1255.00,227.40)(4.504,0.526){7}{\rule{6.257pt}{0.127pt}}
\multiput(1255.00,224.34)(40.013,7.000){2}{\rule{3.129pt}{0.800pt}}
\multiput(1308.00,234.39)(5.709,0.536){5}{\rule{7.267pt}{0.129pt}}
\multiput(1308.00,231.34)(37.918,6.000){2}{\rule{3.633pt}{0.800pt}}
\put(305,113){\raisebox{-.8pt}{\makebox(0,0){$\ast$}}}
\put(358,129){\raisebox{-.8pt}{\makebox(0,0){$\ast$}}}
\put(411,144){\raisebox{-.8pt}{\makebox(0,0){$\ast$}}}
\put(463,157){\raisebox{-.8pt}{\makebox(0,0){$\ast$}}}
\put(516,165){\raisebox{-.8pt}{\makebox(0,0){$\ast$}}}
\put(569,169){\raisebox{-.8pt}{\makebox(0,0){$\ast$}}}
\put(622,172){\raisebox{-.8pt}{\makebox(0,0){$\ast$}}}
\put(675,173){\raisebox{-.8pt}{\makebox(0,0){$\ast$}}}
\put(727,175){\raisebox{-.8pt}{\makebox(0,0){$\ast$}}}
\put(780,177){\raisebox{-.8pt}{\makebox(0,0){$\ast$}}}
\put(833,180){\raisebox{-.8pt}{\makebox(0,0){$\ast$}}}
\put(886,184){\raisebox{-.8pt}{\makebox(0,0){$\ast$}}}
\put(939,189){\raisebox{-.8pt}{\makebox(0,0){$\ast$}}}
\put(991,195){\raisebox{-.8pt}{\makebox(0,0){$\ast$}}}
\put(1044,201){\raisebox{-.8pt}{\makebox(0,0){$\ast$}}}
\put(1097,207){\raisebox{-.8pt}{\makebox(0,0){$\ast$}}}
\put(1150,213){\raisebox{-.8pt}{\makebox(0,0){$\ast$}}}
\put(1203,220){\raisebox{-.8pt}{\makebox(0,0){$\ast$}}}
\put(1255,226){\raisebox{-.8pt}{\makebox(0,0){$\ast$}}}
\put(1308,233){\raisebox{-.8pt}{\makebox(0,0){$\ast$}}}
\put(1361,239){\raisebox{-.8pt}{\makebox(0,0){$\ast$}}}
\end{picture}
\end{center}
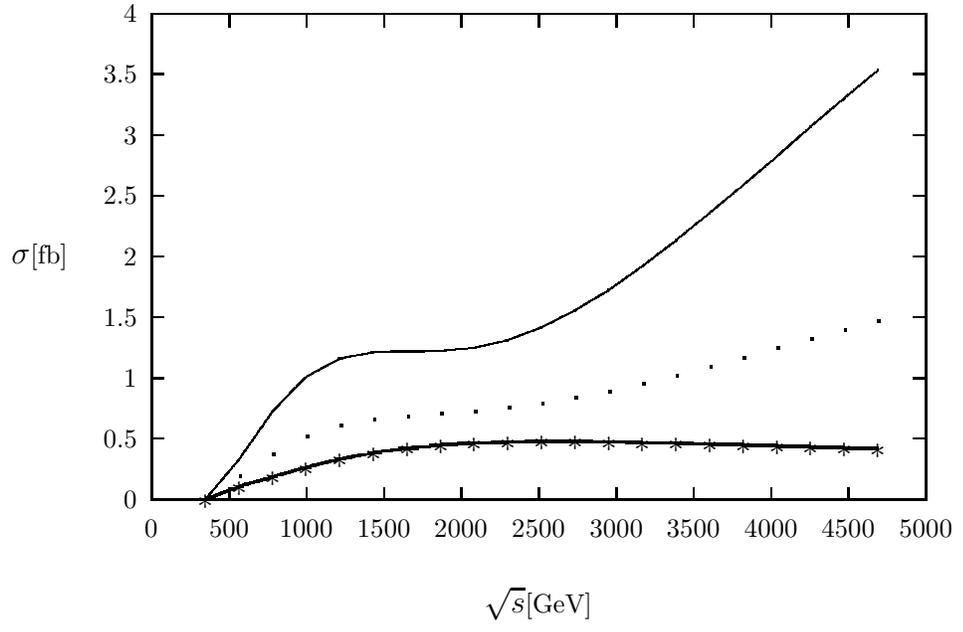
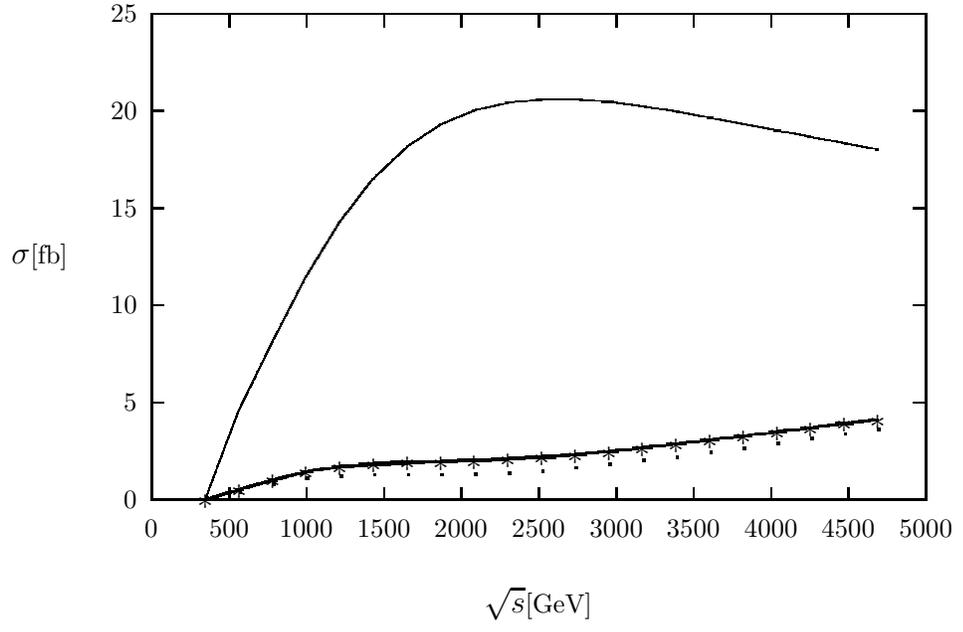
\end{document}